# Innovation of Superparamagnetism in Lead Nanoparticles

Thirugnanasambandan Theivasanthi[*] and Marimuthu Alagar

Centre for Research and Post Graduate Department of Physics, Ayya Nadar Janaki Ammal College, Sivakasi - 626124, Tamilnadu, India.

*Corresponding author: Phone: +91-9245175532   E-mail: theivasanthi@pacrpoly.org

___

**Abstract:** *This work reports, some of the features related to the room temperature superparamagnetic behaviours of lead nanoparticles. Such behaviours have not been discussed elaborately, in any literatures, so far. It is obvious from our studies that the superparamagnetic behaviours of Pb nanoparticles are surprising behaviours, since bulk Pb has diamagnetic behaviours. Hysteresis curve from Vibrating Sample Magnetometer study of Pb nanopowder (metal) serves some new observations. The successful calculations of values like magnetic size, saturation magnetization, coercive field, and remanent magnetism confirm its superparamagnetic nature and explore its potential application in various industries. This work throws some light on and helps further research.*

___

## 1. INTRODUCTION

*SuperParaMagnetic* (SPM) behaviours describe, the magnetisation (M) as a function of magnetic field (B) for a population of randomly oriented, non-interacting, single domain particles, whose magnetisation remains in thermal equilibrium with the applied field. The individual magnetic moments of all atoms in nanoparticles create the total magnetic moment of the nanoparticles. This total moment is regarded as a giant magnetic moment. Pavel *et al*. describe the phenomenon of magnetic viscosity is known as SPM effect [1].

It is a general and well known fact that nano-materials are behaving differently from their bulk material [2]. Size and shape control many of the physical properties (viz., melting point, magnetism, specific heat, conductivity, band gap, etc.), luminescence, and optical, chemical and catalytic properties of nanomaterials [3]. Based on these facts, the present research has been done, to observe the SPM behaviours of metallic lead (Pb) nanoparticles. *Vibrating Sample Magnetometer* (VSM) study of Pb nanopowder confirms its SPM behaviour. The results disclose the absence of hysteresis loop, negligible values of coercivity and remanence of the sample material.

The use of nanotechnology has shown a high impact on society and environment. Scientists show high interest for magnetic nanomaterials due to its potential applications. Also bulk materials show different magnetic behaviours, when prepared in nanoscale [4]. In comparison to macroscopic objects, nanoparticles have very high 'surface to volume' ratio and quantum mechanical effects. Therefore, the magnetization of nanoparticles is dominated by finite size and surface effects [5-6]. The crystallinity also affects the magnetic properties [7].

Magnetic materials having around a dozen nanometers in size are showing SPM behaviors in which coercivity and remanence are zero value. Nanoparticles (with special properties like SPM) are important material for applying to magnetic targeting carriers [8]. Magnetic nanoparticles have been thoroughly studied during the last decades due to their many promising applications in chemical, physical and medical fields [9]. They are utilised in various technological applications such as data storage devices, where every particle represents a bit of information [10].

Sundaresan *et al*. concluded that metal oxides such as $CeO_2$, $Al_2O_3$, ZnO, $In_2O_3$, and $SnO_2$ exhibit diamagnetism but their nanoparticles exhibit room-temperature ferromagnetism [11]. Magnetic nanoparticles exhibit a variety of unique magnetic phenomena that are drastically different from their bulk counterparts. The fundamental magnetic properties like coercivity (Hc) and susceptibility ($\chi$) of these nanoparticles depends on various factors such as size, shape, and composition [12].

$Fe_3O_4$ nanoparticles exhibit SPM behaviours at room temperature due to large fraction of surface atoms in the nanocrystals but bulk Fe3O4 shows ferrimagnetic behaviour [13-16]. Specifically, the domain wall structure of the bulk crystalline ferrites is replaced by a single domain structure of each particle, which leads to new phenomena like SPM [17]. The particles of magnetite below 10–15 nm size contain only one single magnetic domain [18]. It is known that when the particle size is smaller than 30 nm, magnetite and maghemite particles display SPM properties, *i.e.*, they are attracted to a magnetic field but retain no magnetism after removal of magnetic field [19]. In the absence of an external field, the net magnetization of SPM materials at room temperature is zero [20].

While, the particle size of magnetite particles decreases below 30 or 20 nm, their saturation magnetization (*Ms*) value also decreases due to newly arisen phenomena of SPM behaviours [21-22]. The *Ms* Value of SPM Iron Oxide nanoparticles is smaller than the bulk Iron Oxide sample due to nonmagnetic layer on the surface of SPM Iron Oxide nanoparticles [23]. The absence of hysteresis loop, coercivity and remanent magnetization at room temperature indicate that the nanoparticles exhibit SPM behaviour [24]. It is known that *Ms* of single-domain SPM nanoparticles is size-dependent [25] and it is explained by the small surface effect [26].

In order to switch the magnetization into a different state, a certain energy barrier needs to be overcome. If this energy originates from thermal energy, such particles are called SPM. There are no longer stable magnetization configurations but the magnetic moment permanently switches between different orientations. SPM particles show no hysteresis loop and their magnetization response to an external magnetic field resembles the Langevin behaviour of paramagnetic materials [27]. Soft magnetic materials with particle size below 10 nm show SPM (switching) properties and these materials are prospective for many applications in biomedical and its related field. The applications of these nanomaterials are depending on various factors like size of the particles, stability, ambient magnetic properties, and biocompatibility [28] SPM materials play an important role in biomedical applications including magnetic resonance imaging for clinical diagnosis, magnetic drug targeting, hyperthermia anti-cancer strategy [29-31].

We have made an attempt to find the room temperature, SPM behaviours of spherical shaped Pb nanoparticles. It is a surprise that bulk Pb has diamagnetic behaviours. In this study, we will present some of the new findings by magnetic characterization of Pb nanoparticles synthesized by electrolysis using a bioactive compound - konjac aqueous extract. These findings suggest that the synthesized material is an efficient SPM material and can be utilized for various industrial purpose like Magnetic Inks, Magnetic separation, Vacuum sealing, Magnetic marking, Magnetic refrigeration, MRI, Magnetic Data Storage and Research tools in materials physics, geology, biology and medicine. In conclusion, we have shown that nanoparticles behave differently from bulk. To our knowledge, such advanced insights have so far not been said for Pb nanoparticles.

## 2. EXPERIMENTAL

In order to explore the magnetic behaviours of lead nanopowder, we have synthesized it, in accordance with our (T.Theivasanthi and M.Alagar) earlier literature procedure [32]. TEM analysis from the same literature confirms the FCC structure, 10 nm average size, and spherical shape of Pb nanoparticles. X-Ray Diffraction (XRD), Energy Dispersive X-Ray (EDAX), Differential Scanning Calorimetry (DSC), Atomic Absorption Spectroscopy (AAS), Fourier Transform-Infra Red (FT-IR) and theoretical density calculation studies of the same report confirms that the sample was Pb metal Nanoparticles. VSM analysis of this present study also confirms the nano nature of the sample i.e. particle diameter is 44 nm.

5 g of $Pb(NO_3)_2$ salt was dissolved in 100 mL of distilled water and transferred to an electrochemical bath (volume: $4 \times 3 \times 3$ cm$^3$) had lead rod as working electrode (anode), stainless steel rod as counter electrode (cathode). A constant voltage of 15 V was applied between the electrodes using a power supply for 10 minutes. At the end of the process, deposition of Pb nanoparticles was observed and they were removed from both cathode and bath (Particles which were settled down on the electrolytic cell). 20 g of konjac tuber was sliced into many pieces and boiled for 10 minutes with 100 mL of distilled water. At the end, konjac aqueous extract was decanted, few drops of extract was added to the synthesized Pb nanoparticles (to prevent oxidation & stabilization) and were kept in a hot air oven at 50 °C for two hours / until it dried.

VSM (Model: Lakeshore VSM 7410) was used to characterize the DC magnetic properties of the prepared sample of Pb spherical nanoparticles. The powder sample was placed in a glass ampoule (measuring cap) mounted in VSM and was fixed in paraffin in order to exclude the motion of powder. Moment Vs field measurement for 44 mg (0·044 g) mass of the sample was taken at room temperature 303 K and applied Field Range ± 20000 G. When the sample was placed in that magnetic field $B$, a magnetic moment 'm' or magnetization '$M$' (magnetic moment per unit volume) was induced which related to $B$. The magnetization variations were studied and hysteresis parameters such as saturation magnetization (M$s$), coercivity (*Hci*) and remnant magnetism (*Mr*) were measured.

## 3. RESULTS & DISCUSSIONS

VSM is a very common and versatile method of measuring magnetic properties which determines the magnetic moment by vibrating the sample perpendicular to a uniform magnetic field in between a set of pickup coils [33]. In a VSM, a sample is placed suitably within sensing coils, and is made to undergo sinusoidal motion, i.e., mechanically vibrated. The resulting magnetic flux changes induce a voltage in the sensing coils that is proportional to the magnetic moment of the sample.

VSM results showed that SPM properties of spherical lead nanoparticles. A plot of typical magnetic moment Vs magnetic field hysteresis curve traced at room temperature for the lead nanoparticles is shown in fig.1, where the magnetization hysteresis loops appeared S-like or sigmoid shape and extremely thin due to no remanence and negligible coercivity. It was showed a very steep initial rise in magnetization with varying applied magnetic field and then a smooth change of magnetization with more gradual increase to saturation.

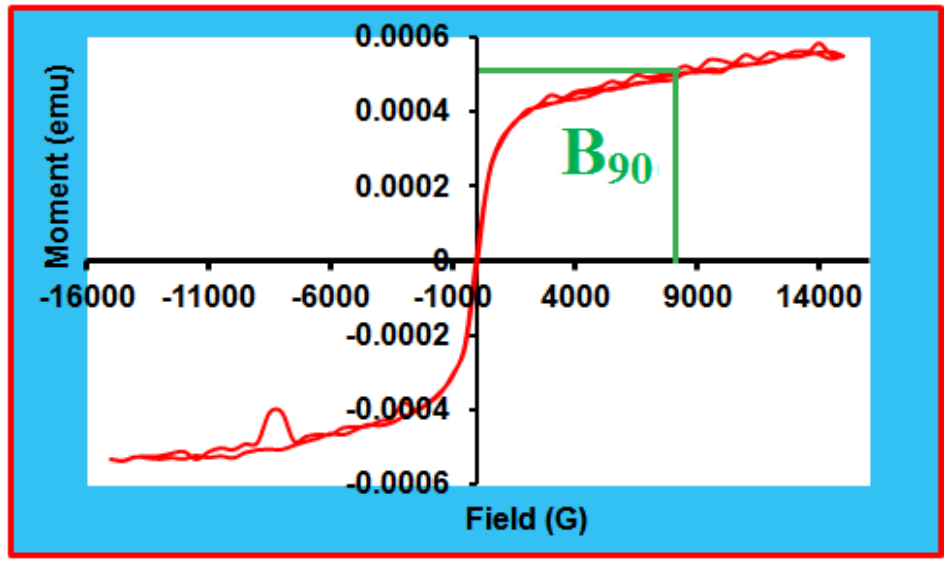

**Figure.1.** Hysteresis Curve of Lead Nanoparticles.

No appreciable hysteresis, negligible width of the loop, negligible coercivity and the no saturated magnetization are indicating SPM behaviour. This change is due to quantum size effects and increased surface area of nanosized magnetite particles which create SPM [34]. When the descending curve (the position of the loop traced between Bmax and 0) and the ascending curve (between 0 and Bmax) are the same, there are no magnetic hysteresis and no magnetic remanence. The treatment is reversible. Paramagnetic, diamagnetic and SPM behaviour are all reversible. The field necessary to reduce the net moment to zero is defined as the coercive field ($\mu_o H_c$) or coercivity. Saturisation remanence (Mr) is the y intercept of descending curve. Bulk coercivity (Bc) is the x intercept of ascending loop. Jackson suggested that the y intercept pointed to a cubic anisotropy [35].

The experimental hysteresis gives the maximum field values of descending and ascending (15000.4 G and -0.349051 G) and (-0.000999883 G and 15000.4 G) curves respectively. It does not exhibit magnetic hysteresis loop, coercivity (Hc) and remanent magnetization (Mr) because of very small or negligible values. Jung *et al.* explain that remanence to saturation ratio (Mr/Ms) characterizes the squareness of the hysteresis loops [36]. Mr/Ms << 0.01 and Hr/Hc > 10 are the typical values for SPM particles. The squareness ratio (SQR) i.e. Mr/Ms of the sample is 0.03517 which indicates no squareness. It appears in sigmoid or S-like shape, without any loop which proves that spherical lead nanoparticles are SPM in nature. The net magnetization in the absence of an external field is zero. The related measurement values are given in Table. 1.

**Table.1.** Magnetic Measurements Values of Lead Nanoparticles.

| Coercivity (Hci) G | Retentivity (Mr) emu | Magnetization (Ms) emu | Field at Ms G | Sensitivity emu |
|---|---|---|---|---|
| 39.269 | 0.000019718 | 0.00056059 | | |
| Negative -52.239 | Negative -0.000013156 | Negative -0.00053986 | 14000 | -6.2000 |
| Positive 26.299 | Positive 0.00002628 | Positive 0.00058132 | | |
| Mass: 44 mg. Number of Points: 151. Time Constant: 1 sec. Elapsed Time: 27 min 53 sec ||||| 

Morrish *et al.* have reported that $Fe_3O_4$ nanoparticles having particle size ~ 50 nm or even less may be considered as a single domain [37]. The reason for SPM nature of the lead nanoparticles is essentially due to the very small size. The smaller size may be considered as equivalent to a single magnetic domain where the thermal vibrations easily overcomes energy barrier resulting SPM nature of nano size particles.

$B_{90}$ is a field, at which the magnetization reaches 90% of saturation, on room temperature. It also occurs in the field at which the SPM particles population reaches 90% saturation. It is a quick guide to the SPM slope (the SPM susceptibility $\chi_{sp}$) contributing to the hysteresis response. It is very sensitive to particle size with very steep slopes and for the particles at the SPM / Single Domain threshold. Tauxe *et al.* argue that the threshold size is ~ 20 nm [38]. The initial susceptibility $\chi_0$ was determined by fitting a straight line to the M-H data [39]. A straight line fit with high field to find the susceptibility of sample is shown in Fig.2.

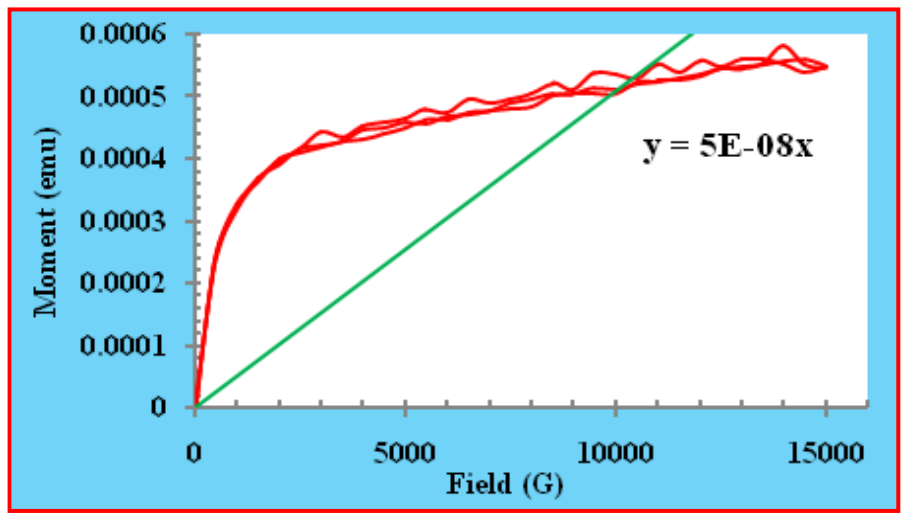

**Figure.2.** High field susceptibility of Lead Nanoparticles.

The hysteresis curve of the sample yields a small or negligible amount of *Ms* and coercivity of the magnetic phase. If any magnetic phase presents, the *Ms* Value will yield its amount. A small or negligible amount of magnetic phase may be present resulting in an initial susceptibility value.

In literatures, it is said that *Ms* value of SPM magnetite nanoparticles is smaller than the bulk sample due to nonmagnetic layer on the surface of SPM nanoparticles but in this present experiment it is not alike and *Ms* value is more than bulk. Magnetization (magnetic moment per unit volume) *M* of nano Pb is differ from bulk lead. Being a diamagnetic substance, bulk lead has no unpaired electron orbital or spin angular momentum, negative molar susceptibility value ($\chi_m$) and it varies only slightly with temperature. The specific magnetization of pure Pb is also negative. Some magnetic characters of bulk and nano Pb are enumerated in Table.2.

**Table.2.** Magnetic characters differences between Bulk and Nano Lead.

| Characters | Nano Lead | Bulk Lead |
|---|---|---|
| **In the presence of an external magnetic field** | | |
| Magnetism | Superparamagnetism | Diamgnetism |
| Susceptibility χ | $5*10^{-8}$ (less than ferromagnet and more than paramagnet) | Molar $\chi_m$ = -23*$10^{-6}$ $cm^3$ $mol^{-1}$ <br> Mass χ = $-1.5 \times 10^{-9}$ <br> Volume χ = $-1.8 \times 10^{-5}$ <br> Magnetic χ = $-0.12 \times 10^{-6}$ <br> (at temperature 7.18 K) |
| Relative permeability | 1.00000005 | less than unity |
| Ms | 0.00056059 emu | < 0 (negative) |
| Mr/Ms ratio | 0.03517 | - |
| Average moment | 0.00047 emu | Very small / Negligible |
| **In the absence of an external magnetic field** | | |
| Magnetization | 0 | 0 |
| Magnetic moments | 0 | 0 |
| **Temperature >0K** | | |
| Magnetic moment | 0 | - |

### 3.1. Magnetic Moment

The experimental magnetic moment is calculated from the following formula [40].

$$\eta = \frac{[MW * Ms]}{5585} \quad \quad \quad \quad (1)$$

Where $M_W$ is molecular weight of the sample and Ms is saturation magnetization in emu/g.

### 3.2. Particle Size Calculation

When the size of magnetic nanoparticles reduces below a critical value (13nm for magnetite), such magnetic anisotropy increases the thermal energy ($K_BT$) more than the energy barrier. Hence, the $K_BT$ is able to reorient the domains, diminishes hysteresis and coercive field to zero (which are characteristic features of SPM regime). When the particle size of a magnetic phase is reduced, the coercivity increases, goes through a maximum, and then tends towards zero [41]. Domains reorientation is shown in Fig.3.

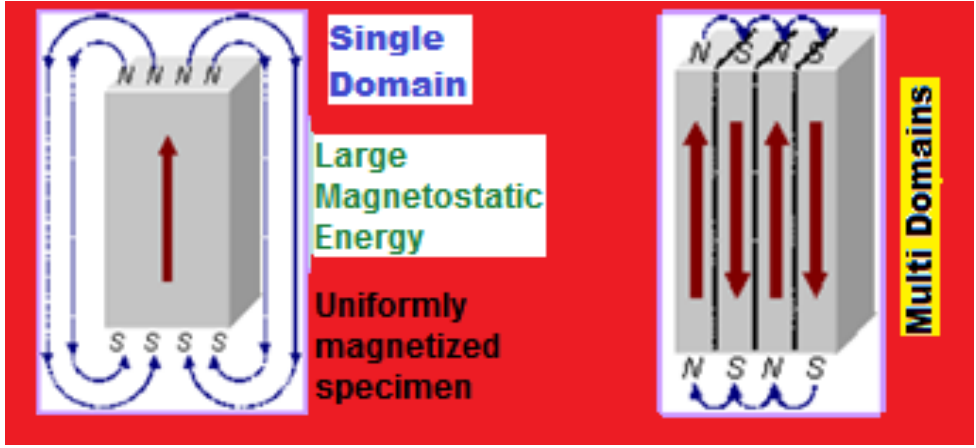

**Figure.3.** Schematic diagram of domain reorientation. Break up of magnetization into domains due to energy minimization. Giant magnetic moment of SPM Regime in SD.

Hysteresis data are used to characterize the domain state and infer the average magnetic grain size. Bean *et al.* proposed firstly to determine the size distribution from the measurements of magnetization curve based on the theory of SPM [42]. In many cases, an accurate knowledge on the average size and size distribution of nanoparticles are crucially important for the proper working of the particular application. The particle moment may be determined by magnetic measurements, from which the particle volume may be obtained as *m*0/*Ms,* if *Ms* is known [43]. One can estimate $K_{eff}$ particle volume and particle size (diameter) by using the following relations [44-45].

$$K_{eff} = \frac{H_c M_S}{2} \quad \text{................................. (2)}$$

$$V = \frac{25 K_B T_B}{K_A} \quad \text{....................... (3)}$$

$$V = \frac{4}{3}\pi r^3 \quad \text{................................. (4)}$$

Where, $K_{eff}$ and $K_A$ are effective magnetic anisotropy constant, $H_C$ is coercivity, $M_S$ is saturation magnetization, $K_B$ is Boltzmann constant (1.38 *$10^{-23}$), $T_B$ is blocking temperature (303 K), V is particle volume and r is radius of particle. We estimated the values that $K_{eff}$ = 2502 J.m$^{-3}$, particle volume = 4.2*$10^{-23}$ m$^3$, radius = 2.15*$10^{-8}$ m and particle size (diameter) = 44 nm for lead SPM nanoparticles. The transition temperature from a magnetism state to SPM state with no hysteresis behaviour is referred as $T_B$. Bulk materials have magnetic anisotropic energies much larger than the thermal energy (kT).

## 4. CONCLUSIONS

First time, we have made an attempt to find the SPM behaviours of Pb nanoparticles (metal) and such advanced insights have so far not been said in literatures. We have successfully calculated its particle size (diameter) as 44 nm, some other values such as saturation magnetization, coercive field, and remanent magnetism from VSM analysis. It is confirmed from our experiment that lead nanoparticles do not exhibit diamagnetism as like bulk lead, exhibit SPM behaviours and the results are discussed in detail. The results related to the SPM suggest, the synthesised lead nanoparticles for a lot of potential applications in various industries. It is concluded that the present work is a clear evidence for the SPM nature of Pb spherical nanoparticles, at room-temperature. This work throws some light on and helps further research on nano-sized lead powder.

## Acknowledgements

The authors express their immense thanks to Department of Science and Technology (DST) and SAIF, IIT Madras, Chennai, for providing VSM instrument to analyze the sample. They also acknowledge assistances and encouragements of staff & management of PACR Polytechnic College, Rajapalayam and Ayya Nadar Janaki Ammal College, Sivakasi, India.